\documentclass[11pt]{article}
\pdfoutput=1 % if your are submitting a pdflatex (i.e. if you have
   
\usepackage{jheppub} % for details on the use of the package, please
\usepackage{url}
\usepackage{caption}
\usepackage{subcaption}

\usepackage{times}

\usepackage[latin9]{inputenc}
\usepackage{float}
\usepackage{amsmath}
\usepackage{amssymb}
\usepackage{dsfont}
\usepackage{graphicx}
\usepackage{esint}
\usepackage{hyperref}
\usepackage{comment}
\usepackage{color}
\usepackage{microtype}
\usepackage{cleveref}
\usepackage{breakurl}
\usepackage{bbm}

\usepackage{mathtools}
\usepackage{enumerate}
\usepackage{tensor}
\usepackage{mathrsfs}
\usepackage{upgreek}
\usepackage{amsbsy}
\usepackage{xcolor}
\hypersetup{
	colorlinks=true,
	linkcolor=blue,
	filecolor=magenta,      
	urlcolor=cyan,
}
\newcommand{\be}{\begin{equation}}
\newcommand{\ee}{\end{equation}}
\newcommand{\ben}{\begin{displaymath}}
\newcommand{\een}{\end{displaymath}}
\newcommand{\bea}{\begin{eqnarray}}
\newcommand{\eea}{\end{eqnarray}}

   \newcommand{\rf}[1]{(\ref{#1})}

\def\be{\begin{equation}}
\def\ee{\end{equation}}
\def\bea{\begin{eqnarray}}
\def\eea{\end{eqnarray}}
\def\bit{\begin{itemize}}
\def\eit{\end{itemize}}

 \makeatletter
\usepackage{mathtools}
\usepackage{enumerate}
\usepackage{tensor}
\usepackage{mathrsfs}
\usepackage{upgreek}
\usepackage{amsbsy}
\usepackage{xcolor}
\hypersetup{
	colorlinks=true,
	linkcolor=blue,
	filecolor=magenta,      
	urlcolor=cyan,
}
\allowdisplaybreaks

\makeatother

\makeatletter
\DeclareRobustCommand{\rcite}[1]{%
  \rcite@aux#1,\@nil{#1}%
}
\def\rcite@aux#1,#2\@nil#3{%
  \if\relax#2\relax
    Ref.~\cite{#3}%
  \else
    Refs.~\cite{#3}%
  \fi
}
\makeatother

\hypersetup{
    colorlinks = true,
    citecolor = {blue},
    linkcolor = {blue},
    urlcolor = {blue},
}

\title{\rm {\LARGE \bf  Quantization of Gravity in Spherical Harmonic Basis}}

\author{Renata Kallosh }

\affiliation{Stanford Institute for Theoretical Physics and Department of Physics,\\ Stanford University, Stanford, CA 94305, USA}

\emailAdd{kallosh@stanford.edu}

\notoc

\abstract{ We perform  canonical quantization of gravity in the background of a Schwarzschild black hole in the generalized Regge-Wheeler gauge proposed in \cite{Kallosh:2021ors}. We find that 
 the Hamiltonian at the quadratic level is unitary and ghost-free.   Two  canonical degrees of freedom are associated with  Zerilli-Moncrief  and Cunningham-Price-Moncrief functions of the metric perturbations. The ${\ell }<2$ part of the Hamiltonian vanishes. This quantization with the unitary Hamiltonian  for  gravity is valid also in Minkowski space in spherical coordinates.
}

\begin{document}

\maketitle

\newpage

\tableofcontents{}

%\newpage

\section{Introduction}

This note is a development based on a recent paper \cite{Kallosh:2021ors} where we have performed a covariant (Lagrangian) quantization of gravity in a black hole background in the Regge-Wheeler  set up \cite{Regge:1957td,Zerilli:1971wd,Martel:2005ir}.  The gauge-fixing condition in  \cite{Kallosh:2021ors} includes the Regge-Wheeler gauge for ${\ell }\geq 2$ modes, and a certain background covariant gauge  for ${\ell }<2$  modes, where Regge-Wheeler gauge is not valid. We will refer to the  gauge in  \cite{Kallosh:2021ors} covering all ${\ell }$ modes, as a `generalized  Regge-Wheeler gauge'.

The Feynman path integral for gravity, viewed as quantum field theory (QFT), is  defined by De Witt-Faddeev-Popov  \cite{DeWitt:1967ub,Faddeev:1967fc} and takes a form, in absence of sources 
\be
 \int  D h
 \,J_\chi ( g,h)  \delta \big (\chi_\alpha ( g,h)\big ) e^{\mathrm{i} S ( g+h) } \, .
\label{pi}\ee
Here we integrate over the perturbations $h$ in the background metric $g$. The gauge-fixing conditions are $\chi_\alpha(g, h)=0$.  The Jacobian designed to make this path integral independent on the choice of the gauge-fixing conditions   can be presented with the help of the Faddeev-Popov (FP) ghosts \cite{Faddeev:1967fc} \be
J_\chi = \int D\bar C^\alpha DC_\beta \, e^{\mathrm{i} \int \mathrm{d}^4 x \, \bar C^\alpha(x)  Q_{\alpha} {}^{ \beta}  (g, h)C_\beta(x) }\, .
\ee
The differential operator in the ghost action is defined by the gauge variation of the gauge-fixing finctions
$
\delta \chi_\alpha = Q_{\alpha}{}^{\beta} \xi_\beta
$. For the  choice of the gauge-fixing functions made in \cite{Kallosh:2021ors}, which in addition to Regge-Wheeler gauge for $l\geq 2$ modes includes gauges for $l<2$ modes, a generalized  Regge-Wheeler gauge, we have found that the ghost actions do not have time derivatives  in  Schwarzschild coordinates. We therefore predicted that in the a generalized  Regge-Wheeler gauge  \cite{Kallosh:2021ors} the canonical Hamiltonian according to the rules for gauge theories 
\cite{Faddeev:1969su,Fradkin:1970pn,Faddeev:1973zb} is expected to be unitary.

In this note we will  present the quadratic in gravitational perturbations $h$ part of the gravity Hamiltonian in the spherical harmonic basis. Before doing this we will perform the standard counting of physical degrees of freedom in this case. The structure of the Hamiltonian will confirm this counting.

The standard counting of physical degrees of freedom in gauge theories in the QFT context of the Feynman path integral is the same in either Lagrangian or Hamiltonian quantization, and it is also gauge independent, if performed correctly. 
The general counting formula is formulated for the number of gauge field components equal to $n+k$ in case of $k$ gauge symmetries.  The total number of physical degrees of freedom is
\be  n-k
\label{number}\ee
This final counting formula in QFT is valid for any choice of  gauge-fixing, but  the procedure is different for unitary and pseudo-unitary gauges. 
For example in 4D the metric has $n+k=10$   components and there are $k=4$ gauge symmetries, the counting is $n-k=(10- 4)-4=2$.

In QFT
in the class of unitary gauges Hamiltonians have manifestly ghost-free underlying Hilbert spaces. There are $(p^*, q^*)$ variables in Faddeev's theorem \cite{Faddeev:1969su} as described in \cite{Kallosh:2021ors}.
This means that all $n-k$ physical states have positive definite metric. The S-matrix is unitary.
\be
\# \, {\rm degrees \, of\, freedom}_{\rm unitary \, H} \,  =  n-k
\label{numberU}\ee
Meanwhile, 
in other gauges, for example, 4D Lorentz covariant gauges in gravity,  the Hamiltonians are ``pseudo-unitary'' with underlying state spaces with negative-norm ghost degrees of freedom \cite{Fradkin:1977hw,Batalin:1977pb}. In such case the counting goes as follows: there are $n+k$ states with positive norm and $2k$ states with negative norm presented by FP anti-commuting ghosts, so the total counting, with account of negative norm states,  is the same as in unitary gauges
\be
\# \, {\rm degrees \, of\, freedom}_{\rm pseudo-unitary \, H}  =  n+k -2k  \Rightarrow n-k
\ee
 The S-matrix is pseudo-unitary in a space of states with the indefinite metric.

We will see that the quadratic in $h$ part of the gravity Hamiltonian in spherical harmonic basis does support  this counting. In the class of gauges used in  \cite{Kallosh:2021ors} the canonical Hamiltonian is unitary, as predicted there.

The complete form of the Hamiltonian to all orders of $h$  is beyond the scope of this paper. However, in \cite{Kallosh:2021ors} we have argued that the non-linear couplings of ghosts to all orders in $h$ are free of time derivatives on ghosts. Therefore one would expect that the non-linear in $h$ terms in $H$ will be consistent with the unitarity of the Hamiltonian which will be deduced in this note  at the level quadratic in $h$.

 The corresponding part of the action $S( g+h)$, quadratic in $h$, is of the form
\be
S= {1\over 2} \int h_{\mu\nu} S^{ \mu\nu  \lambda \delta} (g) h_{\lambda \delta}
\label{quad} \ee
Here $S^{ \mu\nu  \lambda \delta} (g)$ is a differential operator depending on the background metric $g$. The left hand side of equations of motion ${\delta S\over \delta h_{\mu\nu} }=0$ linear in $h$ takes the form
\be
Q^{\mu\nu} \equiv {\delta S\over \delta h_{\mu\nu} }= S^{ \mu\nu  \lambda \delta} (g) h_{\lambda \delta}
\label{EOM}\ee
One can restore the action in eq. \rf{quad} from the information available in eq. \rf{EOM}.

In the spherical harmonic basis the 4D spacetime  is split into $\mathcal{M} = \mathcal{M}_2\times\mathbb{S}^2$  with coordinates $(x^{a},\theta^A)$, $a=1,2$.  The 4D perturbations $h_{\mu\nu} $ are  represented by  2D fields for each $({\ell }, m)$ \cite{Regge:1957td,Zerilli:1971wd,Martel:2005ir}. 
The corresponding equations are known and we will use them as derived in \cite{Martel:2005ir} in Schwarzschild coordinates. Once the quadratic Lagrangian is known, it is possible to derive the relevant quadratic in $h$ Hamiltonian.
For ${\ell }\geq 2$ modes the corresponding quadratic Hamiltonian was constructed by Moncrief in \cite{Moncrief:1974am} where also the relevant Regge-Wheeler \cite{Regge:1957td} and Zerilli \cite{Zerilli:1971wd} equations were re-derived in the form of Hamiltonian equations of motion. The Hamiltonian was derived in \cite{Moncrief:1974am} in absence of source terms.
For ${\ell }<2$ the Hamiltonian was not studied, to the best of our   knowledge. In \cite{Moncrief:1974am} it was explained that  the  attention was restricted to modes with $l\geq 2$ since the modes with $l<2$ are nonradiative and require a special treatment.

Here we will use the known field equations \rf{EOM} in the form given in  \cite{Martel:2005ir} in Schwarzschild coordinates, which allow to derive the Lagrangian in \rf{quad}. From the quadratic Lagrangian we derive  a canonical  quadratic part of the Hamiltonian, with account of the algebraic constraints in our gauges.  We will conclude that there are no physical degrees of freedom suitable for quantization  at ${\ell }<2$. Our definition of quantized degrees of freedom involves the QFT quantization conditions in 2D space of the form
\be
[q(r, t) , p(r', t) ]= i \delta (r-r')
\label{quant}\ee
The classical field equations for low multipoles  in presence of sources are known to have non-trivial solutions. For example for the monopoles ${\ell }=m=0$   there are solutions  like 
$
h^{00}_{tt} \sim {\delta M\over r}
$, 
they are known to affect  the  the black hole mass. However,  there are no solutions of the constraint equations compatible with the quantization condition \rf{quant} for ${\ell }<2$.

All our results are valid for any mass $M$ of the Schwarzschild black hole, and the limit to $M = 0$ is continuous. This means that they apply not only to the quantization in the black hole background, but also to the unitary quantization of the gravitational field in the Minkowski space background in spherical coordinates.

\section{ Counting Gravity  Physical Degrees of Freedom in the Spherical Harmonic Basis}
The ansatz of Regge-Wheeler for the metric perturbations $h_{\mu\nu}$  with  spherical harmonics of definite parity  is given in  \cite{Regge:1957td,Zerilli:1971wd,Martel:2005ir}. In our 
 recent paper \cite{Kallosh:2021ors} it was adapted for the purpose of quantization following  the formalism and notations in \cite{Martel:2005ir}. In particular, we have presented the gauge symmetry transformations to all orders in $h$. The background metric in Schwarzschild coordinates is
 \begin{equation}
g_{\mu\nu}dx^\mu dx^\nu=	-f(r)\,\mathrm{d}t^2 + \frac{\mathrm{d}r^2}{f(r)} + r^2(x)\,\mathrm{d}\Omega^2_2\, ,  \qquad f(r)  = 1-\frac{2GM}{r}
	\label{SchldBackgroundGauge}
\end{equation} 

\

 The 2D fields representing all components of $h_{\mu\nu}$ in 4D include the following  
\bea \label{RWA}
&& h_{ab}^{\ell m(+)}, \quad j_{a}^{\ell m(+)}, \quad K^{\ell m(+)}, \quad G^{\ell m(+)}\,  \hskip 3 cm {\ell } >1  , \quad  {\rm even}\\
%\cr
&&  h_{a}^{\ell m(-)}, \quad h_2^{\ell m(-)}\,    \hskip 6.5 cm {\ell } > 1 , \quad  {\rm odd} \\
%\cr
&&h_{ab}^{1 m(+)}, \quad  j_{a}^{1 m(+)},  \quad K^{1m(+)}   \qquad \hskip 4 cm {\ell }=1, \quad  {\rm even}
\label{ansatzDe} \\
%\cr
&&h_{a}^{1 m(-)}  \hskip 8.2 cm  {\ell }=1, \quad   {\rm odd}\\
%\cr
&&h_{ab}^{0 0(+)},  \quad K^{00(+)}  \hskip 6.6 cm  {\ell }=0,  \quad  {\rm even}
\label{ansatzM} 
 \eea
The gauge symmetries are also expanded in spherical harmonics. In the form   given in our recent paper \cite{Kallosh:2021ors} these are
 \bea
\xi^{{\ell } > 1 }  \quad  {\rm even}  \qquad   &&\Rightarrow \qquad \{ \xi_{a}^{\ell m(+)}, \xi^{\ell m(+)} \}
\label{sym}\\
%\cr
\xi^{{\ell } > 1}  \quad  {\rm odd}   \qquad   &&\Rightarrow \qquad \{  \xi^{\ell m(-)} \}
\label{sym}\\
%\cr
\xi^{{\ell }=1}  \quad  {\rm even}  \qquad   &&\Rightarrow \qquad  \{ \xi_{a}^{1 m(+)}, \xi^{1 m(+)} \} 
\label{symDe} \\
%\cr
\xi^{{\ell }=1}  \quad  {\rm odd}  \qquad  && \Rightarrow \qquad  \{  \xi^{1 m(-)}  \} \\
%\cr
\xi^{{\ell }=0}   \quad  {\rm even}   \qquad   &&\Rightarrow \qquad  \{  \xi_{a}^{0 0(+)} \}
\label{symM} 
 \eea
The gauge symmetry parameters $\xi_{a}^{\ell m (+)}$, $\xi^{\ell m (+)}$, $\xi^{\ell m (-)}$  
can be regarded as scalar and vector fields on $\mathcal{M}_2$.

The counting of physical degrees of freedom in these 5 sectors is
\begin{enumerate}
  \item ${\ell } >1   \quad  {\rm even}:  \quad  \, n+k = 7, \quad k= 3\, \quad \Rightarrow \quad n+k -2k = 7- 2\cdot 3=1$
  \item ${\ell } >1   \, \, \quad  {\rm odd}:  \quad  \, n+k = 3, \quad k= 1\, \quad \Rightarrow \quad n+k -2k= 3- 2\cdot 1=1$
  \item ${\ell } =1    \quad  {\rm even}:  \quad  \, n+k = 6, \quad k= 3\, \quad \Rightarrow \quad n+k -2k= 6- 2\cdot 3=0$
  \item ${\ell } =1   \, \,  \quad  {\rm odd}:  \quad  \, n+k = 2, \quad k= 1\, \quad \Rightarrow \quad n+k -2k= 2- 2\cdot 1=0$ 
  \item ${\ell } =0    \quad  {\rm even}:  \quad  \, n+k = 4, \quad k= 2\, \quad \Rightarrow \quad n+k -2k= 4- 2\cdot 2=0$
\end{enumerate}
Thus we find that in $l\geq 2$ sector there is one even and one odd physical degree of freedom for each $({\ell },m)$. There are no degrees of freedom for any of ${\ell }<2$.

\section{Quadratic Lagrangian/Hamiltonian  for ${\ell }\geq 2 $ Modes}
\subsection{${\ell }\geq 2 $ even} 
 There are 7 fields here, $h_{ab}^{\ell m(+)}, \quad j_{a}^{\ell m(+)}, \quad K^{\ell m(+)}, \quad G^{\ell m(+)}$.
 There are 7 equations of motion for these fields.   Now we can add the 3 Regge-Wheeler gauge-fixing conditions
 \be
 G=j_a=0
 \ee 
The remaining 4 fields are $h_{ab}^{\ell m(+)},  K^{\ell m(+)}$. We expect to identify 3 constraints which will leave us with just one canonical degree of freedom. These equations are according to \cite{Martel:2005ir}
 \begin{eqnarray} \label{MarP}
Q^{tt} &=& -\frac{\partial^2}{\partial r^2} {K} 
- \frac{3r-5M}{r^2 f} \frac{\partial}{\partial r} {K} 
+ \frac{f}{r} \frac{\partial}{\partial r} {h}_{rr}
+ \frac{(\lambda+2)r + 4M}{2r^3} {h}_{rr} 
+ \frac{\mu}{2r^2 f} {K}, \\ 
Q^{tr} &=& \frac{\partial^2}{\partial t \partial r} {K} 
+ \frac{r-3M}{r^2 f} \frac{\partial}{\partial t} {K} 
- \frac{f}{r} \frac{\partial}{\partial t} {h}_{rr} 
- \frac{\lambda}{2r^2} {h}_{tr}, \cr
Q^{rr} &=& -\frac{\partial^2}{\partial t^2} {K} 
+ \frac{(r-M)f}{r^2} \frac{\partial}{\partial r} {K} 
+ \frac{2f}{r} \frac{\partial}{\partial t} {h}_{tr} 
- \frac{f}{r} \frac{\partial}{\partial r} {h}_{tt}  
+ \frac{\lambda r + 4M}{2r^3} {h}_{tt} 
- \frac{f^2}{r^2} {h}_{rr} 
- \frac{\mu f}{2r^2} {K}, \cr 
Q^\flat &=& -\frac{\partial^2}{\partial t^2} {h}_{rr} 
+ 2 \frac{\partial^2}{\partial t \partial r} {h}_{tr} 
- \frac{\partial^2}{\partial r^2} \tilde{h}_{tt} 
- \frac{1}{f} \frac{\partial^2}{\partial t^2} {K} 
+ f \frac{\partial^2}{\partial r^2} \tilde{K} 
+ \frac{2(r-M)}{r^2 f} \frac{\partial}{\partial t} {h}_{tr} 
- \frac{r-3M}{r^2 f} \frac{\partial}{\partial r} {h}_{tt} 
\cr & & \mbox{} 
- \frac{(r-M)f}{r^2} \frac{\partial}{\partial r} {h}_{rr} 
+ \frac{2(r-M)}{r^2} \frac{\partial}{\partial r} {K} 
+ \frac{\lambda r^2-2(2+\lambda)Mr+4M^2}{2r^4 f^2}{h}_{tt}   
- \frac{\lambda r^2-2\mu Mr-4M^2}{2r^4} {h}_{rr}\nonumber 
\end{eqnarray} 
Here 
\be
\lambda = {\ell }({\ell }+1) \qquad \mu = ({\ell }-1)({\ell }+2)
\ee
The quadratic in $h$ Lagrangian can be restored from these  equations  as explained in eqs. \rf{quad}, \rf{EOM}. One can proceed by defining for each of the 4  fields their canonical momenta. For example, there is no time derivative on $h_{tt}$ in the action, therefore $p_{tt}=0$, the other 3 coordinates in the form of ${\cal L} (q, \dot q)$ do have time derivatives, however, two more combinations of $q$'s and $p$'s are constrained. Only one independent canonical degree of freedom out of 4  is left.

The Hamiltonian of the related system starting with the  Arnowitt, Deser, Misner construction  was derived in \cite{Moncrief:1974am}. We skip the details of the derivation here starting with the field equations 
\rf{MarP} since the answer for the corresponding Lagrangian can be also reconstructed from the 
 Zerilli-Moncrief function \cite{Zerilli:1971wd,Moncrief:1974am} which in Regge-Wheeler gauge is
\[
\Psi_{\rm even} = \frac{2r}{{\ell }({\ell }+1)} \biggl[ {K}   
+ \frac{2f}{\Lambda} \biggl( f {h}_{rr} 
- r  {K}_{,r} \biggr) \biggr], \qquad l\geq 2
\]
where $\Lambda = ({\ell }-1)({\ell }+2)  + 6M/r$. The equation of motion in the form of the Zerilli-Moncrief function $\Psi^{lm}_{even}$ as given in \cite{Martel:2005ir} is
\be
(\Box - V_{\rm even})  \Psi_{\rm even}=S_{\rm even}
\label{ZM}\ee
where $\Box= g^{ab} {\cal D}_a {\cal D}_b$ is the Laplacian operator on ${\cal M}_2$, $V_{\rm even}$ depends on $r$ as well as on  $M$ and on $l$,   and $S_{\rm even}$ is the contribution from sources.  We refer to details given in \cite{Martel:2005ir}, where also the relation between Zerilli-Moncrief function and the  original Regge-Wheeler function is explained.
Equation  \rf{ZM} can be derived from the Lagrangian of the form \rf{quad}
\be
{\cal L} = \sum_{{\ell } \geq 2 ,  m} \Big [  {1\over 2} \Psi_{\rm even}( \Box - V_{\rm even})\Psi_{\rm even}  - 
 \Psi_{\rm even} S_{\rm even}\Big ]\ee
 This can be rewritten in the form producing a quadratic part of the Hamiltonian. With $\Psi_{\rm even}\equiv Q_{\rm even}$ and its canonically conjugate $P_{\rm even}$ and,  in absence of sources
\be
 H_{{\ell } \geq 2, \rm even}= {1\over 2} \sum_{{\ell }\geq 2, m} 
 \int \Big [dr  f(  P^{{\ell }, m } )^{2 }_{\rm even}+  f (Q_{, r}^{{\ell }, m })_{\rm even}^2 +   V_{\rm even}
 (Q^{{\ell }, m })_{\rm even}^2\Big ]
\label{Heven} \ee
where
\begin{equation} 
V_{\rm even} = \frac{1}{\Lambda^2} \biggl[ \mu^2 \biggl(
  \frac{\mu+2}{r^2} + \frac{6M}{r^3} \biggr) 
+ \frac{36M^2}{r^4} \biggl(\mu + \frac{2M}{r} \biggr) \biggr] 
\label{4.26}
\end{equation}
This is an example of the Faddeev's theorem  \cite{Faddeev:1969su}, which we described in   \cite{Kallosh:2021ors}, where starting from the original constrained  variables $(p_i, q^i)$  with constraints $\phi^\alpha(p,q)$ one can perform a canonical transformation with $p'_\alpha =\chi_\alpha (p,q) =0$ and  $ q^{'\alpha} = q^{'\alpha} (p^*, q^*)$ 
so that the independent set of canonical variables is $(p^*, q^*)$. In this particular case we find just one set of $(p^*, q^*)$, which are the 
 Zerilli-Moncrief function $\Psi$ of the original variables,  and its canonical conjugate.

\subsection{${\ell }\geq 2 $ odd} 

There are 3 fields in this sector:  $h_{a}^{\ell m(-)}, \quad h_2^{\ell m(-)}$. In the RW gauge
\be
h_2^{\ell m(-)}=0\, .
\ee
Equations of motion for the remaining two fields are
\begin{eqnarray*} 
P^t &=& - \frac{\partial^2}{\partial t \partial r} {h}_r 
+ \frac{\partial^2}{\partial r^2} {h}_t 
- \frac{2}{r} \frac{\partial}{\partial t} {h}_r \
- \frac{\lambda r - 4M}{r^3 f} {h}_t, \\ 
P^r &=& \frac{\partial^2}{\partial t^2} {h}_r 
- \frac{\partial^2}{\partial t \partial r} {h}_t 
+ \frac{2}{r} \frac{\partial}{\partial t} {h}_t 
+ \frac{\mu f}{r^2} {h}_r,
\end{eqnarray*} 
Restoring the quadratic Lagrangian and using partial integration one can identify one  field which enters into Lagrangian without a time derivative, this is $h_t$.  
\be 
{\cal L} = h_t \Big (- \frac{\partial^2}{\partial t \partial r} {h}_r 
+ {1\over 2} \frac{\partial^2}{\partial r^2} {h}_t 
- \frac{2}{r} \frac{\partial}{\partial t} {h}_r \
- {1\over 2} \frac{\lambda r - 4M}{r^3 f} {h}_t \Big ) + {1\over 2} h_r \Big (\frac{\partial^2}{\partial t^2} {h}_r 
+ \frac{\mu f}{r^2} {h}_r\Big ) 
\ee
Thus we find
\bea
&&p_t=0\\
&&p_r= h_{t, r} - h_{r, t} -{2\over r} h_t
\eea
and there is a constraint for $h_t$ algebraically related to $p_r$  
\be
 \Big ( \partial_r + \frac{2}{r}  \Big ) \Big (p_r + h_{t, r} -{2\over r} h_t
\Big ) 
-  {h}_{t,rr} 
+  \frac{\lambda r - 4M}{r^3 f} {h}_t =0
\ee
Therefore there is one independent degree of freedom  $(h_r, p_r)$. These are Faddeev's $(p^*, q^*)$ variables, exactly one set in agreement with the counting give above. One can write the corresponding Hamiltonian $H(h_r, p_r)$ and the field equations.

On the other hand, the Hamiltonian for this system was already derived in \cite{Moncrief:1974am} in  the framework of the Arnowitt, Deser, Misner construction.  The field equations were derived in \cite{Cunningham:1978zfa}, where the corresponding Cunningham-Price-Moncrief function was introduced. In notation of  \cite{Martel:2005ir}, this function is
\be
\Psi_{\rm odd}^{lm}  = \frac{2r}{({\ell }-1) ({\ell }+2) } \biggl( 
 {h}_{t , r} ^{{\ell }m} - 
 {h}_{r,t}^{{\ell }m}  - \frac{2}{r} {h}_t^{{\ell }m}  
\biggr). 
\ee
This function in terms of canonical variables above depends on $(p_r, h_r)$.
As in the even case  discussed  above we are lead to a single field   equation for the Cunningham-Price-Moncrief  function
\be
(\Box - V_{\rm odd})  \Psi_{\rm odd}=S_{\rm odd}
\label{odd}\ee
Here the expressions for $V_{\rm odd}$ and $  S_{\rm odd}$ are given in \cite{Martel:2005ir}, where also 
the relation between Cunningham-Price-Moncrief  function  and the original Regge-Wheeler function is explained.
With $\Psi_{\rm odd}\equiv Q_{\rm odd}$ the Hamiltonian is

 \be
 H_{{\ell } \geq 2, \rm odd}= {1\over 2} \sum_{{\ell }\geq 2, m} 
 \int \Big [dr  f(  P^{{\ell }, m } )^{2 }_{\rm odd}+  f (Q_{, r}^{{\ell }, m })_{\rm odd}^2 +  \Big ( \frac{{\ell }({\ell }+1) }{r^2} - {6M \over r^3}\Big ) 
 (Q^{{\ell }, m })_{\rm odd}^2\Big ]
 \label{Hodd}\ee
  
\section{Quadratic Lagrangian/Hamiltonian  for ${\ell }<2$ Modes}
\subsection{${\ell }=1$ even} 
Our 6 fields are $h_{ab}^{1 m(+)},   j_{a}^{1 m(+)},   K^{1m(+)}$.  We take a gauge-fixing condition \cite{Kallosh:2021ors}
\be
j_{a}^{1 m(+)}= K^{1m(+)}=0
\ee
The remaining filelds $h_{ab}^{1 m(+)}$ in this gauge satisfy the field equations
\begin{eqnarray*} 
Q^{tt} &=& 
 \frac{f}{r} \frac{\partial}{\partial r}{h}_{rr}
+ \frac{2(r + M)}{r^3} {h}_{rr}, \\ 
Q^{tr} &=& 
- \frac{f}{r} \frac{\partial}{\partial t} {h}_{rr} 
- \frac{1}{r^2} {h}_{tr}, \\
Q^{rr} &=& 
 \frac{2f}{r} \frac{\partial}{\partial t} {h}_{tr} 
- \frac{f}{r} \frac{\partial}{\partial r}{h}_{tt}  
+ \frac{ r + 2M}{r^3} {h}_{tt} 
- \frac{f^2}{r^2}{h}_{rr} 
\end{eqnarray*} 
We can therefore reconstruct the Lagrangian of the form \rf{quad} which will produce these equations.
\be
{\cal L} = h_{tt} Q^{tt} +  \Big (\frac{\partial}{\partial t} h_{tr}\Big)   \frac{ 2f}{r}  {h}_{rr} 
-   h_{tr} \frac{1}{2r^2} {h}_{tr} -h_{rr} \frac{f^2}{ 2r^2}{h}_{rr} 
\ee
We now define $q\equiv  h_{tr}, \, p \equiv  \frac{ 2f}{r}  {h}_{rr}$ and $h_{tt} \equiv \lambda$
\be
{\cal L} =  \dot q p + \lambda Q^{tt} (p, \partial_r p)  
-   q^2 \frac{1}{2r^2} -{1\over 8} p^2
\ee
We integrate out the Lagrange multiplier and find
\be
{\cal L} =  \dot q p 
-   q^2 \frac{1}{ 2r^2} -{1\over 8} p^2
\ee
where 
\be
 \frac{f}{r} \frac{\partial}{\partial r}{rp\over 2f}
+ \frac{2(r + M)}{r^3} {rp\over 2f}=0  \qquad \Rightarrow \qquad  p_{,r} + F(r) p=0
\label{Cp}\ee
The algebraic constraint which $p$ has to satisfy contradicts the  commutation relation which have to be imposed for quantization, as shown in eq. \rf{quant}.
There is no solution of the algebraic constraint \rf{Cp} for the canonical momentum $p(t,r)$ which would be consistent with the quantization condition, only $p=0$ is a consistent one. We conclude there that there are no physical degrees of freedom left in this sector,
\be
H_{{\ell }=1, \rm  even} =0
\ee
 This is in agreement with the counting we presented above.

\subsection{${\ell }=1$ odd} 

There are 2 fields: $h_{a}^{1 m(-)}$. We take a gauge-fixing condition $h_{r}^{1 m(-)}=0$ \cite{Kallosh:2021ors}. In this gauge the remaining field equation is
\be
P^t =  \frac{\partial^2}{\partial r^2} {h}_t 
- \frac{2}{r^2 } {h}_t
\ee
The Lagrangian which will generate this equation is
\be
{\cal L}={1\over 2} h_t 
\Big (\frac{\partial^2}{\partial r^2} {h}_t 
- \frac{2}{r^2} {h}_t \Big  )
\ee
There is one field here where the Lagrangian $ {\cal L}(q)$ does not have time derivative of this field, therefore $p={\delta {\cal L}\over \dot h_t}=0 $. There are no canonical variables here and the Hamiltonian vanishes 
\be
H_{{\ell }=1, \rm  odd} =0
\ee
This is in agreement with the counting we presented above.

\subsection{${\ell }=0$ even} 

There are 4 fields here: $h_{ab}^{0 0(+)},  \quad K^{00(+)}$. We take a gauge-fixing conditions $K=h_{tr}=0$ \cite{Kallosh:2021ors}.
The remaining field equations are
\begin{eqnarray*} 
Q^{tt} &=&  \frac{f}{r} \frac{\partial}{\partial r} {h}_{rr}
+ \frac{r + 2M}{r^3} {h}_{rr} 
, \\ 
Q^{rr} &=&  
- \frac{f}{r} \frac{\partial}{\partial r} {h}_{tt}  
+ \frac{ 2M}{r^3} {h}_{tt} 
- \frac{f^2}{r^2} {h}_{rr}
\end{eqnarray*} 
The Lagrangian which will generate these equations is
\be
{\cal L}=h_{tt} \Big (\frac{f}{r} \frac{\partial}{\partial r} {h}_{rr}
+ \frac{(r + 2M)}{r^3} {h}_{rr} \Big ) -{f^2\over 2r^2} h_{rr}^2
\ee
There are 2 fields, $q^1, q^2$, but there are no time derivatives in the Lagrangian, $p_1=p_2=0$, no canonical variables and the Hamiltonian vanishes
\be
H_{{\ell }=0} =0
\ee
 This is again in agreement with the counting we presented above.

\section{A special role of ${\ell }=0,1$ in quantization of gravity}

Is there any relation between the well known fact about the absence of radiation from monopoles and dipoles in gravity and the fact we observed here, that there are no quantum physical degrees of freedom in monopoles and dipoles when gravity is quantized in spherical harmonics basis? The answer is yes, and it has to do with the tensor nature of gravity, so that radiation starts with quadrupoles ${\ell }\geq 2$. 

Regge-Wheeler ansatz for ${\ell }\geq 2$ has 10 functions depending on coordinates of ${\cal M}_2$ listed in eqs. \rf{RWA}-\rf{ansatzM}. Here we show them in the matrix form contracted with spherical functions.
\be
h_{\mu\nu}^{{\ell }>1} =\begin{pmatrix}
 h^{{\ell }m}_{ab} Y^{{\ell }m}   &   &  &  &  {\color {blue}j^{{\ell }m}_a Y_B^{{\ell }m}} \\
\cr  
 {\color {blue} j^{{\ell }m}_a Y_B^{{\ell }m} } &   &  &  & r^2 K^{{\ell }m} \Omega_{AB} Y^{{\ell }m} + {\color {red} G^{{\ell }m} Y_{AB}^{{\ell }m}}\end{pmatrix}^{(+)} + \begin{pmatrix}
0   &   &  &  &  {\color {blue} h^{{\ell }m}_a X_B^{{\ell }m}} \\
\cr  
 {\color {blue} h^{{\ell }m}_a X_B^{{\ell }m}}  &   &  &  & {\color {red} h^{{\ell }m}_2 X_{AB}^{{\ell }m}} \end{pmatrix}^{(-)} \, .
\ee
The number gauge symmetries  in all cases with ${\ell } >0$ is the same since $\xi_\mu$ is a vector
\be
\xi_{\mu}^{{\ell }>0} =\begin{pmatrix}
\xi^{{\ell }m}_{a} Y^{{\ell }m}   \\
\cr  
 {\color {blue}\xi^{{\ell }m} Y_A^{{\ell }m}}  \end{pmatrix}^{(+)}  \, + \begin{pmatrix}
0    \\
\cr  
 {\color {blue}\xi^{{\ell }m}_a X_B^{{\ell }m} }  \end{pmatrix}^{(-)} \, .
\ee
Therefore we find that  instead of 10 fields (even and odd) as for ${\ell }\geq 2$ we have 8 fields (even and odd)  for ${\ell }=1$, no fields  in red
\be
h_{\mu\nu}^{{\ell }=1} =\begin{pmatrix}
 h^{{\ell }m}_{ab} Y^{{\ell }m}   &   &  &  & j^{{\ell }m}_a Y_B^{{\ell }m} \\
\cr  
 j^{{\ell }m}_a Y_B^{{\ell }m}  &   &  &  & r^2 K^{{\ell }m} \Omega_{AB} Y^{{\ell }m} \end{pmatrix}^{(+)} + \begin{pmatrix}
0   &   &  &  &h^{{\ell }m}_a X_B^{{\ell }m} \\
\cr  
h^{{\ell }m}_a X_B^{{\ell }m}  &   &  &  & 0 \end{pmatrix}^{(-)} \, .
\ee
Therefore from 10-2 =8 states we subtract a double set of 4 symmetries, and find no degrees of freedom for ${\ell }=1$ since  8-8=0.

At  ${\ell}=0$  $Y_{AB}^{00}= X_{AB}^{00}=0$, the terms in red are absent, but also $Y_{A}^{00}= X_{A}^{00}=0$, all blue terms are absent. 
\be
h_{\mu\nu}^{{\ell }=0} =\begin{pmatrix}
 h^{{\ell }m}_{ab} Y^{{\ell }m}   &   &  &  & 0 \\
\cr  
0  &   &  &  & r^2 K^{{\ell }m} \Omega_{AB} Y^{{\ell }m} \end{pmatrix}^{(+)} + \begin{pmatrix}
0   &   &  &  &0 \\
\cr  
0  &   &  &  & 0 \end{pmatrix}^{(-)} \, .
\ee
\be
\xi_{\mu}^{{\ell }=0} =\begin{pmatrix}
\xi^{{\ell }m}_{a} Y^{{\ell }m}   \\
\cr  
0  \end{pmatrix}^{(+)}  \, + \begin{pmatrix}
0    \\
\cr  
0 \end{pmatrix}^{(-)} \, .
\ee
We are left with 4 fields and  2 gauge symmetries, there are no degrees of freedom for ${\ell }=0$: 4-4=0.

\section{Quantization of Gravity in Spherical Harmonics Basis in the Flat Background}
The procedure of Lagrangian quantization performed in \cite{Kallosh:2021ors} as well as the values of the unitary quadratic Hamiltonians presented in this paper, have a smooth limit from the Schwarzschild background to a flat one. In Schwarzschild  coordinates this means that the limit $M \rightarrow 0$ is regular.

 In particular,   Zerilli-Moncrief function for ${\ell }\geq 2$ in  Regge-Wheeler gauge in the limit  $M \rightarrow 0$  is 
\be
\Psi_{\rm even}^{{\ell }m} = \frac{2r}{{\ell }({\ell }+1)} \biggl[ {K}   
+ \frac{2}{({\ell }-1)({\ell }+2)} \biggl( {h}_{rr} 
- r  {K}_{,r} \biggr) \biggr], \qquad {\ell }\geq 2
\ee
The Cunningham-Price-Moncrief function is
\be
\Psi_{\rm odd}^{{\ell } m}  = \frac{2r}{({\ell }-1) ({\ell }+2) } \biggl( 
 {h}_{t , r} ^{{\ell }m} - 
 {h}_{r,t}^{{\ell }m}  - \frac{2}{r} {h}_t^{{\ell }m}  
\biggr 
), 
 \qquad {\ell }\geq 2 \ee
 The quadrartic part of the Hamiltonian in both cases  is
 \be
 H_{\rm even/odd}= {1\over 2} \sum_{{\ell }\geq 2, m} 
 \int \Big [dr ( P^{{\ell }, m } )^{2 }_{\rm even/odd}+  (Q_{, r}^{{\ell }, m })_{\rm even/odd}^2 +   \frac{{\ell }({\ell }+1) }{r^2} 
 (Q^{{\ell }, m })_{\rm even/odd}^2\Big ]
\label{HamFlat} \ee
Here $Q_{\rm even/odd} = \Psi _{\rm even/odd}$ and $P_{\rm even/odd}$ is the corresponding canonical conjugate. At the quadratic level these are the only 2 physical states which appear in the unitary Hamiltonian.

The higher order terms in the each of the quantized actions, at the black hole background and in the flat background still have to be constructed.

\section{A comment on Regge-Wheeler and  Teukolsky formalism and gravity waves}

The Cunningham-Price-Moncrief (CPM) master function and the Zerilli-Moncrief (ZM) master function, which were identified here as canonical variables  in the gravity,  appear to play some role also in a more interesting case of the Kerr black holes. Namely, as pointed out in a  review \cite{Pound:2021qin}, there is a relation via Chandrasekhar transformation between these functions and Teukolsky radial function. Note that 
Teukolsky equations for the Weyl tensor components  use the expansion in terms of the spin-weighted spheroidal harmonics. Such and expansion for the metric starts with ${\ell }=2$.

There is also an interesting relation between the metric perturbation far from the source and our canonical variables in the generalized Regge-Wheeler gauge. Namely, according to  \cite{Pound:2021qin} the gravitational wave strain can be determined directly from CPM and ZM functions of the metric. Using the 
 Chandrasekhar transformation between these functions and Teukolsky radial function, and some properties of $\psi_4= C_{n\bar m n\bar m}$ the gravitational strain was given as

\be
r(h_+ -i h_x) =\sum_{{\ell }\geq 2} \sum_{ |m|  \leq {\ell }} {D\over 2} \Big (\Psi_{\rm even}^{{\ell }m}- i \Psi_{\rm even}^{{\ell }m}\Big ) \, \hskip 1 mm   {}_{- 2}  Y_{{\ell }, m } (\theta, \phi)
\ee
where ${}_{- 2}  Y_{{\ell }, m } (\theta, \phi)$ is the the spin-weighted spheroidal harmonic.
That equality holds in the limit $r  \rightarrow  \infty $  (at fixed $u=t-r_*$).
Here the constant
\be
D= \sqrt {({\ell }-1) ({\ell }+1) ({\ell }+1)}
\ee
is the Schwarzschild limit of the constant that appears in the Teukolsky-Starobinsky identities. Clearly, the cases ${\ell }=0,1$ drop from the formula for the gravitational waves. This is in agreement with the  fact established in this paper that these modes have no physical degrees of freedom.

\section{Summary}

In this note we have counted the number of physical quantized degrees of freedom  of Einstein gravity in spherical harmonic basis using the standard formula:  this number is given by $n-k$,  where  $n+k$ is the number of components of gauge fields and the  gauge theory has $k$ gauge symmetries. For example, in 4D the graviton has  $n+k= 6+4= 10$ components and there are $k=4$  gauge symmetries. The  number of physical degrees of freedom is $n-k= (n+k) - 2k= 10-8=2$.

In spherical harmonic basis we have found that for each ${\ell }, m$ in ${\ell } \geq 2$ sector there is one degree of freedom for even parity states and one degree of freedom for odd parity states. In ${\ell }<2$ sector of gravity we have found that there are no physical degrees of freedom.

To construct the Hamiltonian we start with the Regge-Wheeler formulation \cite{Regge:1957td,Zerilli:1971wd,Martel:2005ir} of   Einstein gravity in spherical harmonic basis in the background of a Schwarzschild black hole. The part of the action $S(g+h)$ quadratic in perturbations $h_{\mu\nu}$ in eq. \rf{quad} can be presented in  spherical harmonic basis using the explicit form of equations of motion linear in perturbations, as shown in eq. \rf{EOM}. We take these explicit expressions $Q^{\mu\nu} = {\delta S (g, h) \over \delta h_{\mu\nu}} $, which are linear in $h_{\mu\nu}$, 
from \cite{Martel:2005ir}, and reconstruct  the part of the action $S(g+h)$ quadratic in perturbations $h_{\mu\nu}$. We impose the generalized Regge-Wheeler gauge
 \cite{Kallosh:2021ors}. The action quadratic in fields we take in Schwarzschild  coordinates and proceed with canonical quantization, defining canonical momenta and constraints. 

For ${\ell } \geq 2$ fields the procedure leads to one independent degree of freedom for even and one for odd modes in each case with ${\ell }, m$, in agreement with the counting of physical degrees of freedom. We conclude that up to a canonical transformation such a Hamiltonian is equivalent to the one presented in \cite{Moncrief:1974am} where the corresponding canonical variables are Zerilli-Moncrief function \cite{Zerilli:1971wd,Moncrief:1974am} for even modes and a  Cunningham-Price-Moncrief function \cite{Cunningham:1978zfa} for odd modes. In \cite{Moncrief:1974am} the modes with ${\ell }<2$ were not studied.

We apply our method also for  ${\ell }<2$ modes. In each sector for ${\ell }=1$, even and odd case and for 
${\ell }=0$ we first reproduce the action from the explicit expressions $Q^{\mu\nu} = {\delta S (g, h) \over \delta h_{\mu\nu}} $  linear in $h_{\mu\nu}$. We use the gauge-fixing condition for low multipoles in  \cite{Kallosh:2021ors} and identify the canonical variables and constraints. In each case the conclusion is that there are no independent unconstrained canonical variables suitable for the quantized Hamiltonian. This is again in agreement with the counting of degrees of freedom performed earlier.

The original goal of this investigation was to develop a consistent method of quantization of gravitational field in the background  of a Schwarzschild black hole  \cite{Kallosh:2021ors}.  However, we found that in Schwarzschild  coordinates the limit $M \rightarrow 0$ is regular, and therefore the quantization procedure is valid in the Minkowski background as well. In this paper we found the  Hamiltonian describing unitary evolution of gravitational perturbations in spherical coordinates, which equally well applies to quantization of gravity in Minkowski  background as well as in the Schwarzschild black hole background. The choice of the generalized Regge-Wheeler gauge in \cite{Kallosh:2021ors} where the gravity Hamiltonian is unitary requires to use the spherical harmonic basis for the metric perturbations. This {\it unitary gauge} is a Regge-Wheeler gauge $G^{\ell m(+)}=j_a^{\ell m(+)}=h_2^{\ell m(-)}=0$ for ${\ell } \geq 2$. For ${\ell } =1$ it is $j_{a}^{1 m(+)}= K^{1m(+)}=h_{r}^{1 m(-)}=0$ and for ${\ell } =0$ it is $K^{00}=h_{tr}^{00}=0$.

In this generalized Regge-Wheeler gauge, the quadratic part of the Hamiltonian for ${\ell }<2$ modes is vanishing, whereas for ${\ell }\geq 2$ it is given in eqs. \rf{Heven}, \rf{Hodd} in the black hole background and in eq. \rf{HamFlat} in Minkowski background. 

\section*{Acknowledgement}
I am   grateful to  A. Barvinsky,  E. Coleman, A. Linde,  E. Poisson,  A. Rahman, P. Stamp,  A. Starobinsky, A. Vainshtein,  A. Van Proeyen  and I. Volovich  for stimulating and helpful discussions.  
I am supported by the SITP, by the US National Science Foundation Grant PHY-2014215 and  by the  Simons Foundation Origins of the Universe program (Modern Inflationary Cosmology collaboration).

\

\

\bibliographystyle{JHEP}
\bibliography{lindekalloshrefs}
\end{document}